\documentclass[aps,prl,twocolumn,showpacs,superscriptaddress,groupedaddress]{revtex4-1}
\usepackage{graphicx}
\usepackage{amssymb}
\usepackage{amsmath}
\usepackage{latexsym}
\usepackage{ulem}
\usepackage{color}
\usepackage{dcolumn}
\usepackage{subfigure}
\usepackage[T1]{fontenc}
\usepackage[utf8]{inputenc}
\usepackage[english,frenchb]{babel}
\usepackage[colorlinks=true,linkcolor=blue,citecolor=blue]{hyperref}%

\usepackage{bm}        % for math
\usepackage{amssymb}   % for math
\begin{document}

\title{Universal quantum transport and impurity band super metallicity in self-similar graphene carpets}

\author{G. Bouzerar}
\email[E-mail:]{georges.bouzerar@univ-lyon1.fr}
\affiliation{ CNRS et Universit\'e Claude Bernard Lyon 1, F-69622, LYON, France}
\author{D. Mayou}
\affiliation{Institut N\'eel, CNRS, F-38000, Grenoble, France}                    
\date{\today}
%\clearpage
\selectlanguage{english}
\begin{abstract}
Fractals, a fascinating mathematical concept made popular in the eighties, remained for decades a beautiful scientific curiosity mainly. With the tremendous advances in nanofabrication techniques, such as nanolithography, it has become possible to design self-similar materials with fine structures down to nanometer scale. Here, we investigate  the effects of self similarity on quantum electronic transport in graphene Sierpinski carpets. We find that a gap opens up in the electron spectrum in the middle of which lies a flat band of zeros energy modes. Although these states have a zero velocity, a supermetallic phase is found at the neutrality point. 
For Fermi energy located in the valence/conduction band and in the presence of a small inelastic scattering the system stays metallic and the transport is found strongly anisotropic.

\end{abstract}
\maketitle

The rapid progress in nanofrabrication techniques such as nanolithography, molecular engineering and 3D printing have made it possible to design complex two and three dimensional multi-scale and self-similar materials with fine structures down to nanometer scale \cite{fowlkes,vyatskikh,kempkes,berenschot}. 
Self-similar materials have already been in use in several areas such as fractal antenna \cite{xiu}, photonic crystal waveguides \cite{wen}, or even heat transfer devices \cite{enfield}. Possibilities to design and grow at will this new class of materials open pathways toward the exploration of new exotic physical phenomena that may have remarkable technological spinoffs. The rapidly growing field of cold atoms on optical/artificial lattices also offers a plateform to address these fundamental issues by directly tuning the physical parameters of model Hamiltonians \cite{belopolski,bloch,lewenstein,cooper}. Recent theoretical studies on quantum effects in fractal lattices have been for instance focusing on the Hall effect \cite{vanveen1,vanveen2,fremling}, plasmon confinement \cite{westerhout} and topological phases\cite{brzezinska}. In these studies the host material of the fractal structure is a simple square lattice. In this work, we investigate how transport is affected by the self-similarity of the underlying lattice in one of the most remarkable two dimensional material of the 21st century, Graphene \cite{das-sarma,castro-neto,geim,falko}. 

This one atom thick material holds a large potential in various technological fields. It is a zero gap semiconductors with high mobility at room temperature \cite{morozov,tan}, it displays a huge thermal conductivity, it is extremely flexible, while being stronger than steel and impermeable to gas and liquids. Among its plethora of astonishing physical properties, the  quantum electronic transport is certainly one of the most intriguing and intensely debated \cite{nomura,ando,hwang,pereira,yuan,pereirab,hafner,fan,cresti,trambly,gattenlohner,mucciolo}. Chirality, that results from the bipartite nature of graphene lattice plays a key role in its unconventional electronic transport properties, such as Klein tunnelling \cite{katsnelson,young} or the minimum conductivity at the neutrality point. Effects of disorder such as C vacancies, short and long range on-site potentials, adsorbates or resonant impurities on the one particle spectrum and transport of the massless Dirac fermions has been the main focus of several numerical studies \cite{yuan,fan,cresti,trambly}. However, no clear consensus on the nature of the transport properties at the neutrality point could emerge. Recently, within large scale numerical calculations, it has been shown unambiguously that in the presence of vacancies (up to 1\%) $\sigma(0)$ remains identical to that of the pristine graphene, e.g.  $\sigma_0=\frac{4e^2}{\pi h}$  \cite{mucciolo}.

\begin{figure}[t]\centerline
{\includegraphics[width=0.99\columnwidth,angle=0]{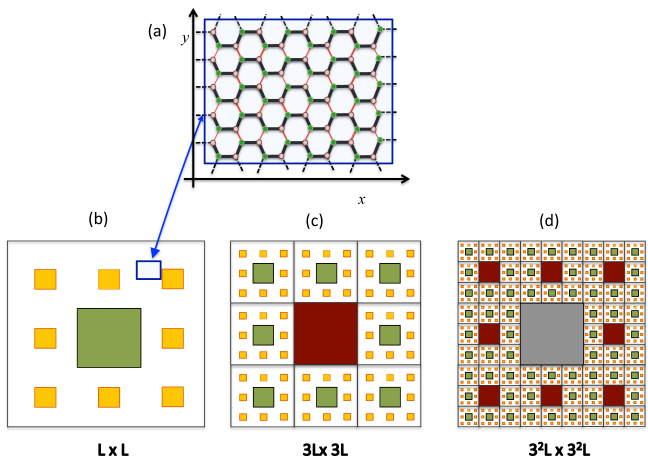}}
\caption{(Color online) Illustration of the GSC, from fractal level 2 to 4. The removed squares produce armchair edges in the $x$-direction and zig-zag in $y$-direction.
}
\label{fig1}
\end{figure}

In this study, the issue of quantum transport in self-similar Graphene Serpinski Carpets (GSC) is addressed. We investigate the interplay between the chirality of the massless Dirac fermions, leading to the unconventional quantum transport in the host compound and the fractal nature of the spectrum resulting from self-similarity. Electrons on GSC's are modelled by a nearest neighbour tight binding Hamiltonian that reads,
\begin{eqnarray}
\widehat{H}=-t \sum_{\left\langle ij\right\rangle,s} c_{is}^{\dagger}c^{}_{js} +h.c.,
\end{eqnarray}
$t=2.7$ eV is the hopping integral, $\left\langle ij\right\rangle$ denotes nearest neighbour pairs of C atoms, c$_{i\sigma}^{\dagger}$ creates an electron with spin s in the $\pi$ orbital at site \textbf{R}$_{i}$. Here, we ignore the next nearest neighbour hopping of the order of 10$\%$ that breaks particle-hole symmetry.
The GSC as illustrated in Fig.~\ref{fig1} are obtained from a square piece of Graphene of size 3$^{i_{c}+1}$a (a is the nearest neighbour C-C distance), on which Sierpinski masks are applied iteratively. In what follows, we make use of the notation (i$_{c}$,f) where L$_{x}$=L$_{y}$=3$^{i_{c}+1}$a is the system size and f the degree of "fractalization", that varies from 0 (pristine) to its maximum value f$_{\text max}$=i$_{c}$. For $f=0$ the system embodies N$_{S}=\frac{4} {3\sqrt{3}a^{2}}$ L$_{x}$.L$_{y}$  C atoms. Periodic boundary along both $x$ and $y$ directions (see Fig.~\ref{fig1}) are used. The smallest system considered corresponds to i$_c$=3 and the largest to i$_c$=7, they contain respectively about 5000 and 33.10$^6$ atoms.
The conductivity along $x$-direction is given by the Kubo formula,
\begin{eqnarray}
\sigma(E)=\frac{e^{2}\hbar}{\pi\Omega} Tr \left[ \operatorname{Im} \widehat{G}_{\eta}(E) 
 \widehat{v}_{x} \operatorname{Im}   \widehat{G}_{\eta}(E)   \widehat{v}_{x} \right]. 
\label{eqcond}
\end{eqnarray}
The current operator defined by $ \widehat{v}_{x} =  -\frac{i}{\hbar}\left[ \widehat{x} ,\widehat{H} \right]$ is, 
\begin{eqnarray}
\widehat{v}_{x}=-i\frac{at}{\hbar}\sum_{i\in A,l,s} \alpha_{l}c_{\textbf{R}_{i}s}^{\dagger} c^{}_{\textbf{R}_{i} + \bm \delta_{l} s} 
 + h.c.
\end{eqnarray}
The sum runs over atoms of A sublattice only, $\bm \delta_{l}$ are the vectors connecting a given atom to its  three nearest neighbours on B sublattice,  and $\alpha_{l}$=1, -$\frac{1}{2}$, -$\frac{1}{2}$  for l=1, 2 and 3 respectively.
The Green's function $\widehat{G}_{\eta}(E)=(E+i\eta-\widehat{H})^{-1}$, $\Omega$ is the Sierpinski carpet area. $\eta$ mimics an energy independent inelastic scattering rate with a characteristic timescale $\tau_{in}= \frac{\hbar}{\eta}$. The calculations are done using the Chebyshev Polynomial Green's function method (CPGF) \cite{mucciolo,weisse} that (i) allows large scale calculations as it requires a modest amount of memory and (ii) a CPU cost that increases linearly with the system size N$_S$. This is in contrast to the exact diagonalization method (ED) that needs a memory scaling as N$^2_S$ and CPU time as N$^3_S$. CPGF has proven to be a powerful tool to address the nature of the magnetic couplings in disordered materials \cite{richard,lee}. In the same spirit as CPGF, the conductivity could be calculated by quantum wave packet dynamics as well \cite{trambly,cresti,triozon}.

\begin{figure}[t]\centerline
{\includegraphics[width=1.10\columnwidth,angle=0]{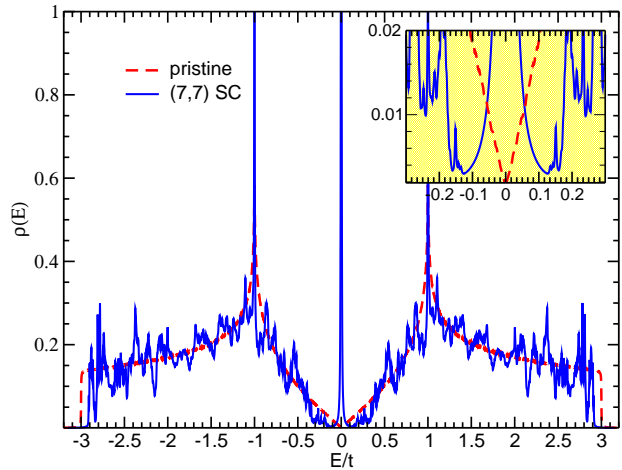}}
%{Fig2-new.eps}}
\caption{(Color online) Density of states of the 7th level Graphene Sierpinski Carpet compared to that of
the pristine case or 0th order fractal level (red dashed).
}
\label{fig2}
\end{figure} 
In Fig.~\ref{fig2} is depicted the electronic density of states (DOS) $\rho(E)=-\frac{1}{\pi N_S} Tr( \operatorname{Im} \widehat{G}_{\eta}(E))$ as a function of energy for the (7,7) GSC. First, we observe a complex fluctuating sub-structures that result from the fractal nature of the eigenspectrum. Sharp peaks are visible at $E=\pm t$, corresponding to the Van Hove singularities in pristine graphene and a third one at $E=0$ that results from the removal of C atoms. This peak corresponds to a flat band of zero energy modes (ZEM). We have indeed checked by ED calculations of smaller systems (see supplementary material) that the eigenvalues are, within numerical accuracy, exactly zero. In the CPGF calculations, from the $E=0$ peak weight, we have extracted the number of ZEM, N$_{zem}$. It coincides with $\vert N_{A}-N_{B}\vert$, N$_{A}$ (resp. N$_{B}$) being the number of atoms on A (resp. B) sublattice, as expected for bipartite lattices \cite{lieb,pereirab}. The density of ZEM modes ($x_{zem}$) found is approximately 0.052.
Interestingly, we find a gap, $\Delta$, between the conduction (resp. valence band) and the flat band as seen more clearly in the inset: $\Delta \approx 0.135~t$, (6,6) and (5,5) carpets give the same value. Thus in GSC's, the ZEM band is an impurity band. This contrasts with the gapless spectrum of randomly distributed vacancies in graphene, unless vacancies are created on the same sublattice \cite{pereirab}. Here, Sierpinski masks produce an unbalance between N$_{A}$ and N$_{B}$, but 2-coordinated C atoms exist on both sublattices. Thus, the gap results only from the self similar structure of the GSC.

\begin{figure}[t]\centerline
{\includegraphics[width=1.0\columnwidth,angle=0]{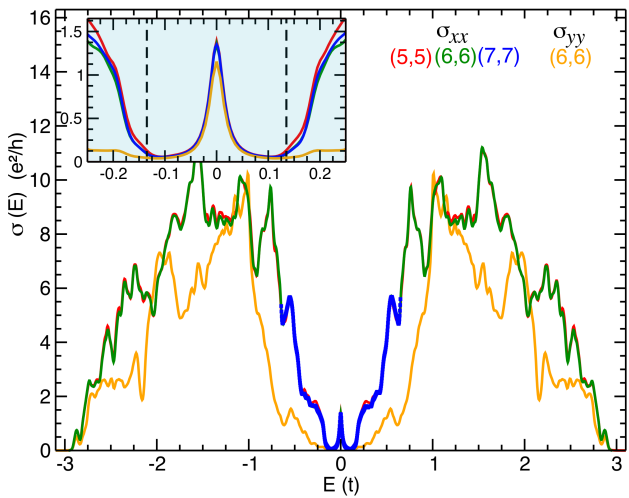}}
\caption{(Color online) 
Conductivity ($x$ and $y$ directions) at T=0~K as a function of the energy. Calculations for (7,7) GSC are done in a restricted region around E=0. The inelastic scattering rate is $\eta=$0.01625t. Along $x$-axis, $\sigma(0)=$1.39, 1.37 and 1.355$\frac{e^2}{h}$ for respectively (5,5), (6,6) and (7,7), and for $y$-direction $\sigma(0)=1.165 \frac{e^2}{h}$ for the (6,6) GSC. The inset magnifies the data around E=0.
}
\label{fig3}
\end{figure}

We now discuss how Sierpinski carpet masks alter the transport. For that purpose, we calculate the dc conductivity $\sigma(E)$ in both $x$ and $y$ directions. Results, for a fixed $\eta$ are depicted in Fig.~\ref{fig3}. The number of random vectors N$_{R}$ used for the trace calculation is 500, 100 and 10 as the system size increases. The number of Chebyshev polynomials kept is M=2000, leading to a $M \times M$ matrix for the moments. It has been checked that both N$_{R}$ and M were sufficient to reach convergence. First, we observe a unexpected anisotropic conductivity in the GSC. Besides restricted regions where they almost coincide, $\sigma_{yy}$ is smaller than $\sigma_{xx}$.  $\sigma_{xx}$ is even 10 times larger than $\sigma_{yy}$  for E in the vicinity of $\pm$0.2 t.
At first glance, it is astonishing that $\sigma_{xx} \ge \sigma_{yy}$ since the edges of the removed square are zig-zag in $y$-direction and armchair in $x$-direction. As it is well known, the nature of the edge in graphene nanoribbons (GNR) has drastic impacts on transport and magnetism \cite{nakada,wakabayashi1,wakabayashi2}. Zigzag edges induce flat band, magnetic moment formation and favour a metallic behaviour whilst armchair GNR are often semiconductors or insulators. Moreover, very small size effects are found, the conductivity for (6,6) and (7,7) GSC's coincide almost with each other. Notice that the carpet size (in units of a) varies from 729 for the (5,5), 2187 for (6,6) and 6561 for the (7,7) GSC. For Fermi energy in the conduction (resp.) valence band and in the presence of a small  $\eta$, the GSC remains metallic for the whole energy range. In addition, due to the fractal nature of the eigenspectrum rich structures and multiple peaks are visible. In the vicinity of $E=0$, $\sigma(E)$ drops
rapidly due to the gap and has a Lorentzian shape. Along the $x$-axis $\sigma(0)$ are 1.39, 1.37 and 1.35 $\frac{e^{2}}{h}$ for respectively the (5,5), (6,6) and (7,7) GSC. In $y$-direction, $\sigma(0)=$1.165$\frac{e^{2}}{h}$ for the (6,6) GSC whilst for (5,5) we have found 1.15 $\frac{e^{2}}{h}$ (not shown).These results indicate a slow convergence towards the universal value $\sigma_{0}$. 
Thus, despite the gap between the ZEM flat impurity band and the valence and conduction bands, the conductivity remains unaltered at E=0. It is worth noticing that for randomly distributed vacancies on the same sublattice (gapped spectrum), $\sigma(E)$ has been found to vanish at this point \cite{cresti}. As discussed in Ref. \cite{mucciolo}, this may result from the difficulty to get converged results due to the singular density of sates. It would be interesting to perform the calculations for this particular gapped case with CPGF \cite{bouzerar}.
\begin{figure}[t]\centerline
{\includegraphics[width=0.95\columnwidth,angle=0]{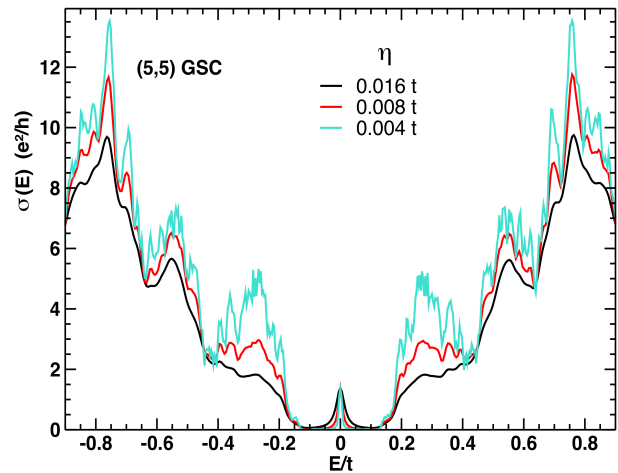}}
\caption{(Color online) 
Conductivity in the $x$-direction at T=0~K as a function of E for three different values of  $\eta$:  $\eta=$ 0.004t, 0.008t and 0.016t.  $\sigma(0) =1.37 \pm 0.02 \frac{e^2}{h}$, thus it is almost insensitive to $\eta$.
}
\label{fig4}
\end{figure} 
In Fig.~\ref{fig4}, for the (5,5) GSC, the effects of varying $\eta$ is illustrated. As $\eta$ reduces, we observe (i) more and more fine structures in $\sigma(E)$, resulting from the fractal nature of the spectrum, and (ii) regions where it increases alternating with narrow energy interval where it remains insensitive. Thus, in the GSC, the conductivity is never Drude like since one would expect an increase proportional to $1/\eta$. This is in contrast with the pristine case (the analytical calculations are straightforward) for which $\sigma(E) \propto 1/\eta$ for $\vert E\vert \gg \eta$. In this case, exactly at the Dirac point, half of $\sigma$(0) originates from inter-band and the other half from the intra-band transitions. In Fig.~\ref{fig4}, $\sigma$(0) is shown to be insensitive to $\eta$ in the GSC. Around E=0, $\sigma$(E) gets narrower and narrower as $\eta$ reduces and can be nicely fitted by a Lorentzian of width $\eta$. For $\eta \ll \Delta$, in the $x$-direction $\sigma(0)$ reduces to,
\begin{eqnarray}
\sigma(0)=\frac{32}{3\sqrt{3}} \frac{e^{2}}{h} \frac{1}{N_{S}}\sum_{\beta} C_{x,\beta},
\end{eqnarray}
where the dimensionless $C_{x,\beta}$ is defined by,
\begin{eqnarray}
C_{x,\beta}= \frac{\hbar^2}{a^2}\sum_{\alpha,\lambda=\pm} \dfrac{\vert \langle  \Psi_{\beta} \vert \widehat{v}_{x}  \vert  \Phi^{\lambda}_{\alpha} \rangle \vert^{2}}{E^{2}_{\alpha}}, 
\label{eqcxbeta}
\end{eqnarray}
$\vert  \Psi_{\beta} \rangle $ are the ZEM eigenstates (E$_{\beta}$=0) and $\vert \Phi^{\lambda}_{\alpha} \rangle $ those of the valence ($\lambda=-$) and conduction ($\lambda=+$) bands respectively with energy $\pm$E$_{\alpha}$ ($\vert$  E$_{\alpha} \vert > \Delta $).
In the limit of vanishing $\eta$ ($\eta \ll \Delta$)  both inter-band and intra-band transitions with matrix elements $\langle \Phi^{\lambda}_{\alpha} \vert \widehat{v}_{x}  \vert \Phi^{\lambda'}_{\alpha'} \rangle$ cannot contribute to $\sigma$(0) because of the gap. On the other hand, the intra-flat band at E=0 can not contribute either because  $\langle  \Psi_{\beta} \vert \widehat{v}_{x}  \vert  \Psi_{\beta'} \rangle$ are all zeros. Indeed, $\vert \Psi_{\beta} \rangle$ and $ \widehat{v}_{x} \vert  \Psi_{\beta'} \rangle$ are orthogonal to each other, they belong to the two different sublattices. Thus, $\sigma$(0) consists only of inter-band transitions between the ZEM impurity band and the valence and conduction bands, a finite conductivity arises despite the presence of the  gap. This feature is rather unusual. In standard systems and E$_{F }$ in the impurity band (localized states), the conductivity is controlled by intra-band transitions only and decays as $\eta$ decreases. However, in quasicrystals and approximants the scenario is different. In icosaedral quasicrystals, the diffusion coefficient is essentially controlled by interband processes which explains the non standard transport  properties observed \cite{quasi1,quasi2,quasi3}.

\begin{figure}[t]\centerline
{\includegraphics[width=0.97\columnwidth,angle=0]{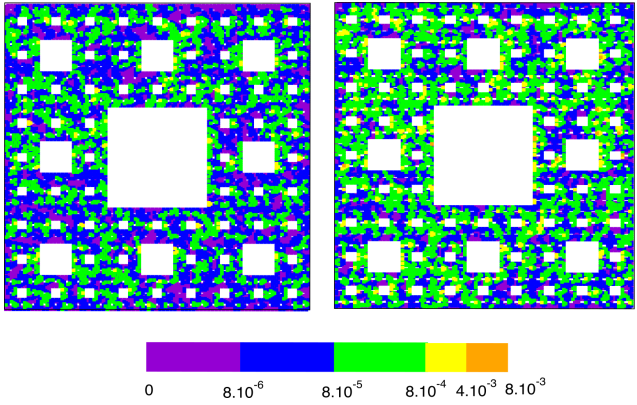}}
\caption{(Color online) 
(left) Local charge density of a typical ZEM state $ \vert \Psi_{\beta} \rangle$  and (right) of $ \widehat{v}_{x} \vert \Psi_{\beta} \rangle$ obtained from exact diagonalization of a (4,4) GSC that contains 26833 C atoms.}
\label{fig5}
\end{figure} 

\begin{figure}[t]\centerline
{\includegraphics[width=0.975\columnwidth,angle=0]{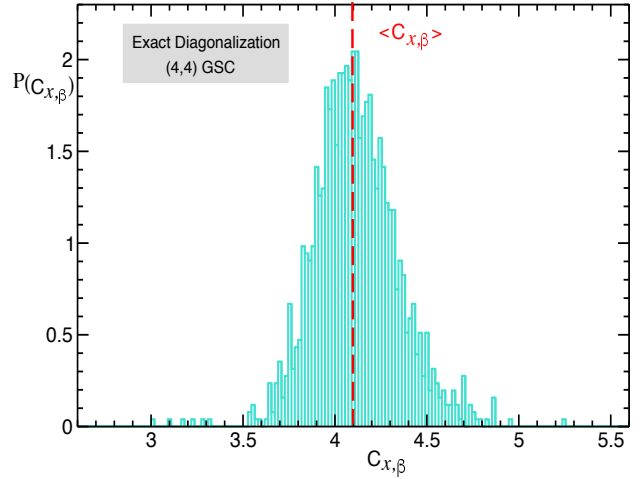}}
\caption{(Color online)Probability distribution of C$_{x,\beta}$, as defined in the manuscript, from ED of (4,4) GSC. $\langle C_{x,\beta} \rangle$ stands for the average value.
}
\label{fig6}
\end{figure}

Lastly, a close look at the local charge distribution of one typical ZEM state $\vert \Psi_{\beta} \rangle$ and of its corresponding $\widehat{v}_{x} \vert \Psi_{\beta} \rangle$ reveals that both are in fact rather extended. This feature is illustrated in Fig.~\ref{fig5} and sheds light on why the overlaps between $\widehat{v}_{x} \vert \Psi_{\beta} \rangle$ and the extended valence/conduction band states $\vert \Phi^{\lambda}_{\alpha} \rangle$ lead to a finite $\sigma(0)$ notwithstanding the existing gap. More quantitatively, the probability distribution of $C_{x,\beta}$'s obtained from exact diagonalization of the (4,4) GSC is plotted in Fig.~\ref{fig6}. This plot displays a relatively narrow distribution with a mean value of $\left\langle C_{x,\beta} \right\rangle = 4.1$ and width of approximately 0.5. From Eq. (4), we immediately obtain $\sigma(0)=1.315 \frac{e^{2}}{h}$ that coincides with $\sigma_{0}$ within less than 3\%. Notice that, from the Einstein formula,
the diffusivity $D(E)$ can be straightforwardly obtained  as well: $D(E)=\frac{e^{2}\rho(E)}{\sigma(E)}$. Here, because of the gap, in the vicinity of E=0, $\rho(E)=\frac{N_{zem} }{\pi\Omega} \frac{\eta}{E^{2}+\eta^{2}}$. From Eq. (\ref{eqcond}) and (\ref{eqcxbeta}) and for $\vert E \vert \ll \Delta $, the diffusivity in the $x$-direction is,
\begin{eqnarray}
D_{x}(E)= \frac{4a^{2}}{\hbar} \left\langle C_{x,\beta}\right\rangle  \, \eta.
\end{eqnarray}
Thus, the diffusivity is proportional to $\eta$ at the neutrality point. This is contrast with the standard $1/\eta$ dependence: $D=\frac{\hbar v^{2}_{F}}{2\eta}$. Because, the diffusivity can be written $D=\frac{L(t)^{2}}{t}$, this allows to extract a typical length-scale $L_{d}=2 \sqrt{ \left\langle C_{x,\beta} \right\rangle}a \approx 4 \,a$ where $\pi L^{2}_{d}$ could be interpreted as the averaged surface occupied by a ZEM eigenstate on the GSC.

In summary, by means of the Chebyshev Polynomial Green's function and exact diagonalization methods, we have investigated the effects of self-similarity on quantum electronic transport in Graphene Sierpinski Carpets. We have found that a finite gap opens up in the electron spectrum in the middle of which lies a flat band of zero energy mode. Although ZEM states have a vanishing velocity, a super-metallic phase is found at E$_{F}$ = 0, the corresponding conductivity is independent of the inelastic scattering rate and coincides within few percent with the universal $\frac{4e^{2}}{\pi h}$. Despite the gap, $\sigma(0)$ originates only from inter-band transitions between the ZEM impurity band and the valence/conduction bands. When E$_{F}$ lies in the valence (resp. conduction) band, and for a small but finite inelastic scattering rate the Sierpinski carpets are found metallic. Lastly, besides the universal character at $E=0$, away the transport is strongly anisotropic.

\begin{acknowledgments}
We would like to thank S. Thébaud for his relevant comments and remarks and P. Mélinon for interesting discussions. 
\end{acknowledgments}

\newpage

\end{document}